\documentclass[preprint,floats,floatfix,aps,12pt]{revtex4}
\usepackage{times} 
\usepackage{amssymb}
\usepackage{amsbsy}
\usepackage{color}

\ifx\pdfoutput\undefined
\usepackage{graphicx}
\else
\usepackage[pdftex]{graphicx,hyperref}
\fi
\graphicspath{{.},{./Fig/}}
\newcommand{\mylab}[1]{\label{#1}}

\newcommand {\dr}{{\mathrm d}\mathbf{r}}

\newcommand {\rr}{\mathbf{r}}

\begin{document}
\title[Modelling the dewetting of evaporating thin films of nanoparticle suspensions]{Modelling approaches to the dewetting of evaporating thin films of nanoparticle suspensions}
\author{U. Thiele\footnote{homepage: http://www.uwethiele.de, u.thiele@lboro.ac.uk}, 
I. Vancea, A. J. Archer, M. J. Robbins, L. Frastia}
\address{Department of Mathematical Sciences, 
Loughborough University, Leicestershire LE11 3TU, UK}
\author{A. Stannard, E. Pauliac-Vaujour, C. P. Martin, M. O. Blunt, P. J. Moriarty}
\address{The School of Physics and Astronomy,
        The University of Nottingham,
        Nottingham NG7 2RD, UK}
\begin{abstract}
  We review recent experiments on dewetting thin films of
  evaporating colloidal nanoparticle suspensions (nanofluids) and
  discuss several theoretical approaches to describe the ongoing
  processes including coupled transport and phase changes. These
  approaches range from microscopic discrete stochastic theories to
  mesoscopic continuous deterministic descriptions. In particular, we
  focus on (i) a microscopic kinetic Monte Carlo model, (ii) a
  dynamical density functional theory and (iii) a hydrodynamic thin
  film model.

  Models (i) and (ii) are employed to discuss the formation of
  polygonal networks, spinodal and branched structures resulting from
  the dewetting of an ultrathin `postcursor film' that remains behind
  a mesoscopic dewetting front. We highlight, in particular, the
  presence of a transverse instability in the evaporative dewetting
  front which results in highly branched fingering structures.  The
  subtle interplay of decomposition in the film and contact line
  motion is discussed.

  Finally, we discuss a simple thin film model (iii) of the
  hydrodynamics on the mesoscale. We employ coupled evolution
  equations for the film thickness profile and mean particle
  concentration. The model is used to discuss the self-pinning and
  de-pinning of a contact line related to the `coffee-stain' effect.

  In the course of the review we discuss the advantages and
  limitations of the different theories, as well as possible future
  developments and extensions.\\[3ex]
%

    \textcolor{red}{\large The paper is published in: 
{\it J. Phys.-Cond. Mat.} {\bf 21}, 264016 (2009), in the Volume ``Nanofluids on solid substrates''
and can be obtained at
    \href{http://dx.doi.org/10.1088/0953-8984/21/26/264016}{http://dx.doi.org/10.1088/0953-8984/21/26/264016}
}
\end{abstract}
\maketitle

\section{Introduction}

%
The patterns formed in dewetting processes have attracted strong
interest since Reiter analysed the process quantitatively in the early
nineties.  In these experiments, that proved to be a paradigm in our
understanding of dewetting, a uniform thin film of
polystyrene (tens of nanometers thick) is deposited on a flat silicon oxide substrate is
brought above the glass transition temperature. The film ruptures in
several places, forming holes which subsequently grow, competing for
space. As a result, a random polygonal network of liquid rims
emerges. The rims may further decay into lines of small drops due to a
Rayleigh-type instability \cite{Reit92,Reit93,ShRe96}. The related problems
of retracting contact lines on partially wetting substrates and the
opening of single holes in rather thick films have also been
studied \cite{deGe85,BrDa89}.

Subsequent work has mainly focused on many different aspects of the dewetting
process for simple non-volatile liquids and polymers  (for reviews
see Refs.\ \cite{Muel03b,Seem05,Thie07}).  All stages of the dewetting of a
film are studied: the initial film rupture via nucleation or a
surface instability (called spinodal dewetting)
\cite{Reit92,Xie98,SHJ01,TVN01,BeNe01,Beck03}, the growth process of individual
holes \cite{RBR91,SHJ01b,Fetz05}, the evolution of the resulting hole pattern
\cite{ShRe96,Beck03}, and the stability of the individual dewetting
fronts \cite{BrRe92,ReSh01,MuWa05}.  We note in passing, that descriptions of
dewetting patterns may also be found in historic papers,
particularly for the dewetting of a liquid film on a liquid substrate.
Tomlinson \cite[footnote 18 on p.\,40]{Toml1870} considered turpentine on
water and Marangoni \cite[p.\,352f]{Mara1871} oil on water.

More recently, interest has turned to the dewetting processes of solutions
and suspensions. However, these systems have not yet been investigated in
any great depth. Such systems are complicated because their
behaviour is determined by the interplay between the various solute
(or colloid) and solvent transport processes. Furthermore, the
solvents that are used often evaporate, i.e., one has to distinguish between `normal'
convective dewetting and evaporative dewetting. A number of
experiments have been performed employing (colloidal) solutions of polymers
\cite{KGMS99,GRDK02,HXL07,LZZZ08}, macromolecules like collagen and
DNA \cite{Mert97,Mert98,TMP98,Maed99,Smal06,ZMCZ08} and nanoparticles
\cite{MMNP00,GeBr00,MTB02,RRGB03,GRPB04,MBM04,Mart07,Stan08,Paul08}.
The latter are sometimes referred to as `nanofluids'.  The initial focus of much of the
research in the field has been on investigating the structures that are formed
which are similar to the ones observed in the
`classical' dewetting of non-volatile liquids.  Labyrinthine
structures and polygonal networks result from spinodal dewetting and
heterogeneous nucleation and growth, respectively.  They are
`decorated' with the solute and therefore conserve the transient
dewetting pattern as a dried-in structure when all the solvent has
evaporated \cite{TMP98,MTB02}. The picture is, however, not
complete. The solute may also shift the spinodal and binodal lines
as compared to the locations of these lines in the phase diagram for the
pure solvent \cite{Vanc08}.  As a consequence, the
solute concentration influences the hole nucleation rate.  More
importantly, the solute particles may also destabilise the dewetting fronts. As a
result, one may find strongly ramified structures in all three systems
\cite{Thie98,GRDK02,Paul08,LZZZ08}. A selection of images exhibiting some
of the possible structures is displayed in Fig.\ref{fig:finger}.

\begin{figure}[htbp]
(a)\includegraphics[width=0.45\hsize]{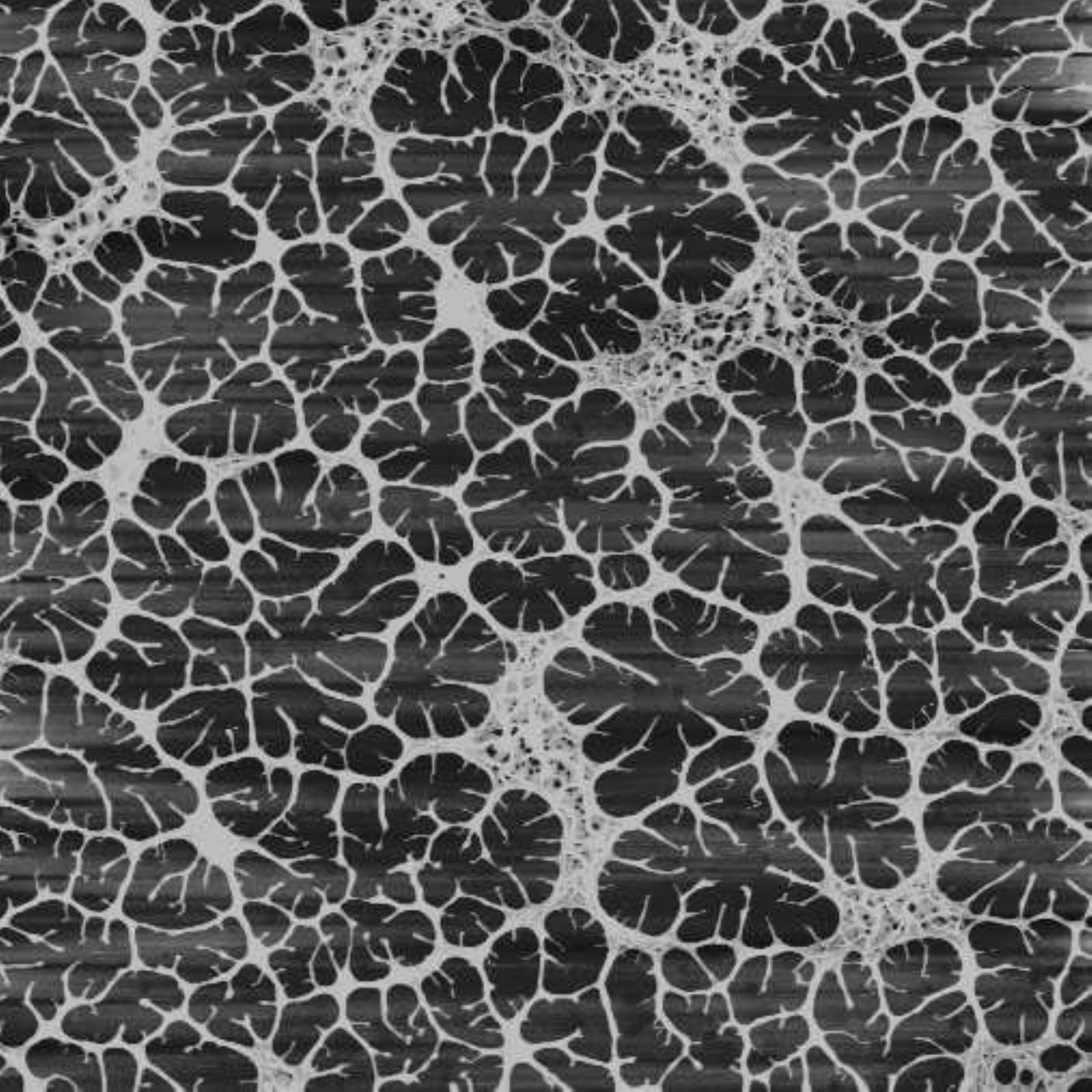}\hspace*{-1.2cm}
\includegraphics[width=0.63\hsize]{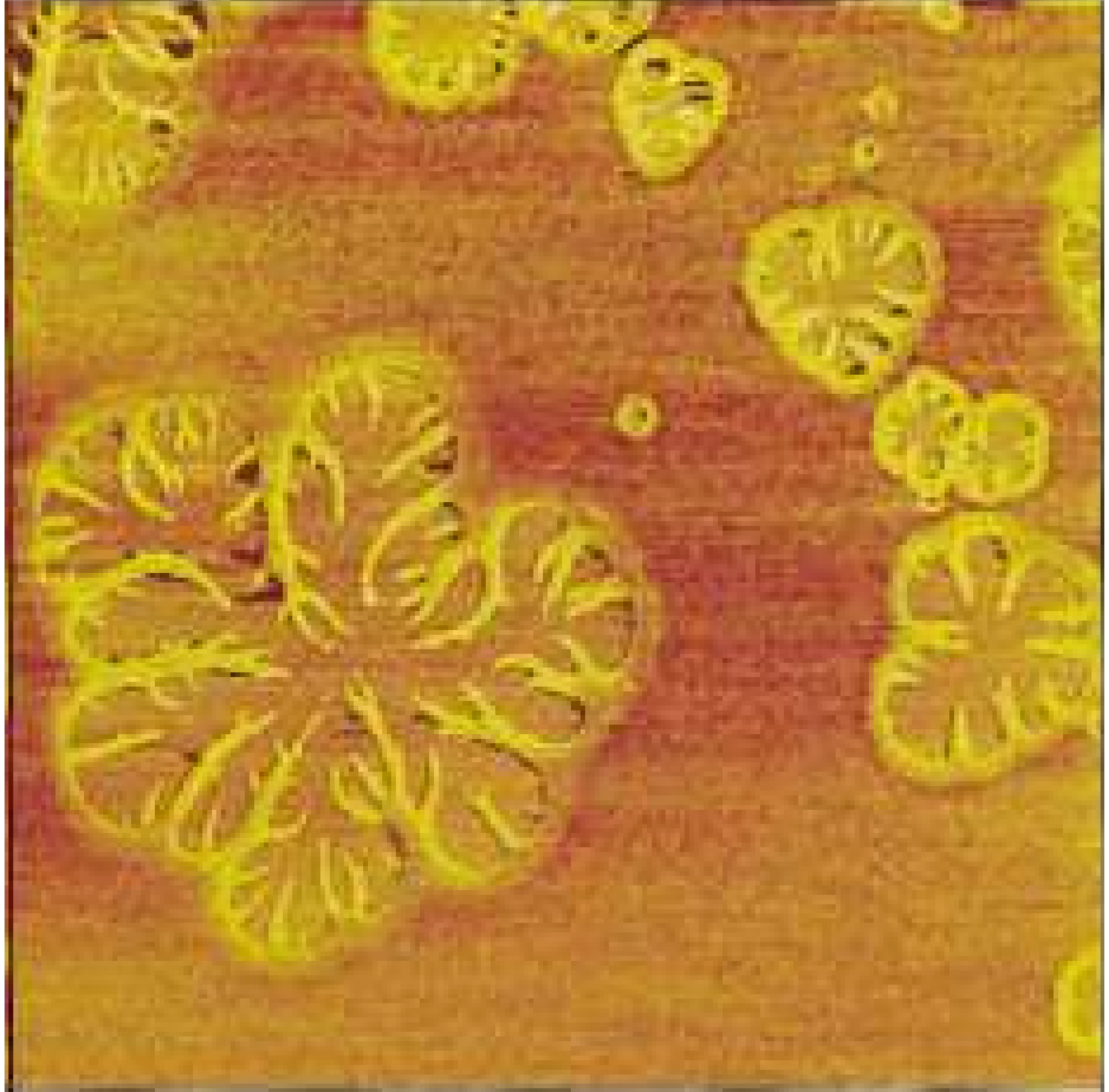}\hspace*{-1.2cm}(b)\\
(c)\includegraphics[width=0.9\hsize]{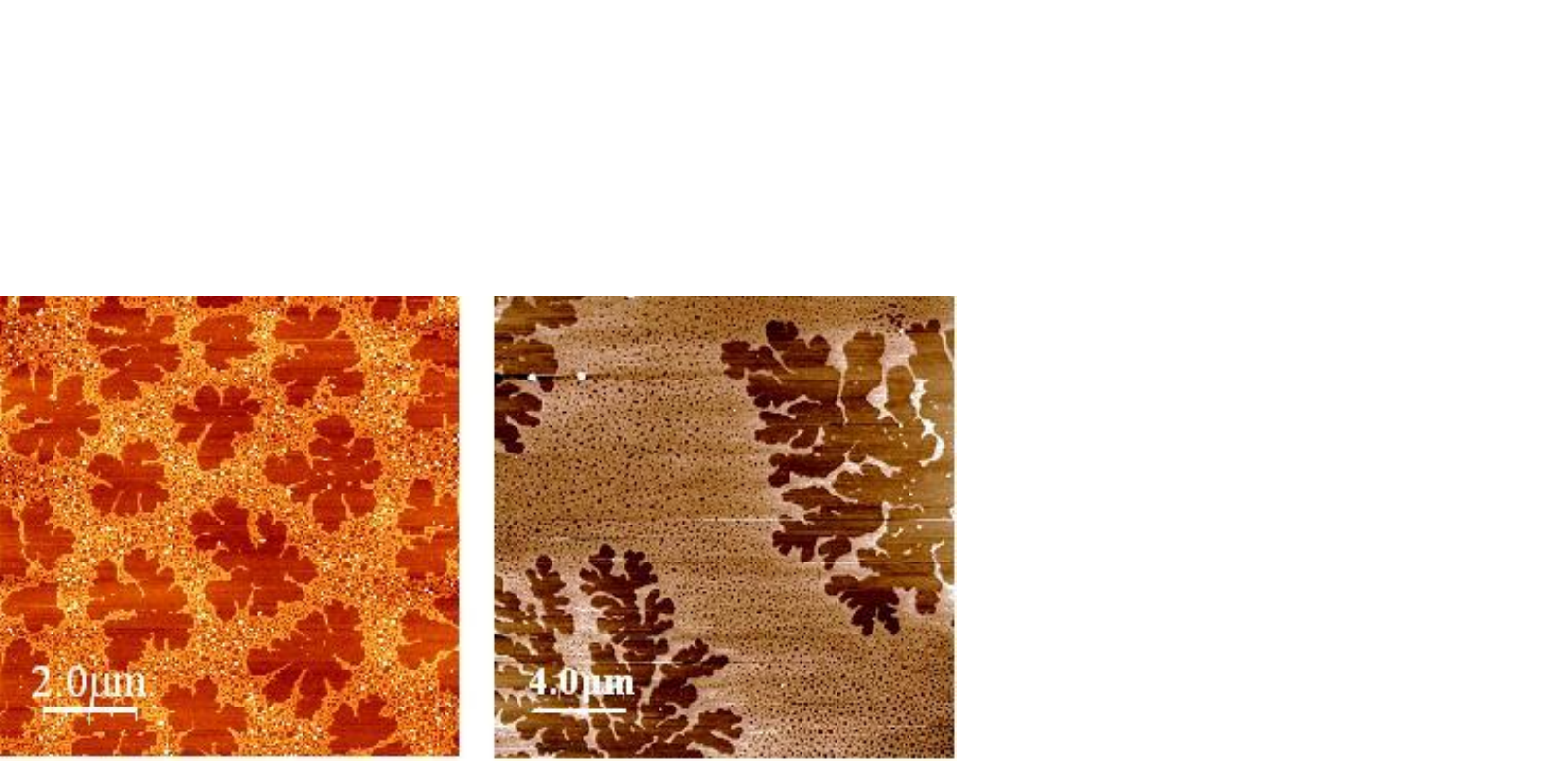}(d)
\caption{(Colour online) Images of strongly ramified dewetting structures obtained
  using Atomic Force Microscopy in the case of (a) an aqueous collagen
  solution on graphite (courtesy of U. Thiele, M. Mertig and W. Pompe;
  see also Ref.\ \cite{Thie98}. Image size: $5\mu$m$\times5\mu$m); (b)
  poly(acrylic acid) in water spin-coated onto a polystyrene substrate
  (reprinted with permission of John Wiley \& Sons, Inc. from Ref.\ \cite{GRDK02}; copyright 
  John Wiley \& Sons, Inc.\ 2002;  Image size:
  $2.5\mu$m$\times2.5\mu$m); and in both
  (c) and (d), a solution of gold nanoparticles in toluene, spin-coated onto
  native oxide terminated silicon substrates (scale bars given in
  panels).  In all the images the lighter areas correspond to the deposited
  solute and the dark areas to the empty substrate.}
\mylab{fig:finger}
\end{figure}

For volatile solvents, the contact lines retract even for wetting
fluids. It has been found that such evaporatively receding contact
lines may deposit very regular line or ring patterns parallel to the
moving contact line \cite{YaSh05,HXL07}. The deposition of a single
ring of colloids from a evaporating drop of colloidal suspension is
well known as the `coffee stain effect' \cite{Deeg97}. Detailed
investigations reveal the emergence of rich structures including
multiple irregular rings, networks, regular droplet patterns, sawtooth
patterns, Sierpinski carpets, and -- in the case of DNA -- liquid
crystalline structures
\cite{ADN95,KGMS99,Deeg00,Deeg00b,SSS02,NgSt02,Smal06}. The deposition
of regularly spaced straight lines orthogonal to the moving contact
line has also been reported \cite{Huan05}. Droplet patterns may as
well be created employing solvent-induced dewetting of glassy polymer
layers below the glass transition temperature
\cite{LYKL04,XSDA07,XSA08}.

Note that the dewetting of pure volatile liquids has also been
studied experimentally \cite{ElLi94} and theoretically
\cite{SLT98,PKS99,LGP02,Pism04}.  In this case, different contact line
instabilities have been observed for evaporating liquid drops
\cite{PBC03,GIMK06}. 

In the present article we review and preview the experiments and in particular
the various modelling approaches for dewetting suspensions of
(nano-)particles in volatile partially wetting solvents. After
reviewing the basic experimental results in Section~\ref{sec:exp}, we
discuss in Section~\ref{sec:theo} several theoretical approaches. In
particular, we present a kinetic Monte Carlo model in
Section~\ref{sec:kmc}, a dynamic density functional theory in
Section~\ref{sec:ddft}, and a thin film evolution equation in
Section~\ref{sec:tfh}. Finally, we conclude in Section~\ref{sec:conc}
by discussing advantages and shortcomings of the individual approaches
and future challenges to all of them.

\section{Experiment with nanoparticle solutions}
\label{sec:exp}

We focus on experiments that use monodisperse colloidal suspensions of
thiol-passivated gold nanoparticles in toluene
\cite{GeBr00,MTB02,MBM04,Mart07,PaMo07,Paul08,Stan08}.  The gold core
of 2 -- 3\,nm diameter is coated by a layer of alkyl-thiol molecules. The
length of the carbon backbone of the thiol used in the experiments
ranges from 6 to 12 carbon
atoms ($C_6$ to $C_{12}$) \cite{Paul08}. By varying the chain length,
one can control to a certain extent the particle-particle
attraction. Normally, the solution is deposited on to a plain silicon
substrate that is covered by the native oxide layer only
\cite{MTB02}. However, one may locally change the wetting behaviour of
the solvent by further oxidising the substrate \cite{Mart07}.  By
adding excess thiol one can also vary the properties of the solvent
\cite{Paul08}.

Two different procedures are employed for the deposition of the
solution on to the substrate: spin-coating or a meniscus technique
\cite{GDD01,PaMo07}. The choice is important as it strongly influences
the evaporation rate and, as a result, the pattern formation
process. When using spin-coating, one finds that directly after
deposition, evaporation competes with dewetting until all the solvent
has evaporated. The resulting deposits of nanoparticles are imaged by
atomic force microscopy (AFM).  For spin-coated films, the evaporation
rate is high and structuring is normally finished before the spin-coater is
stopped. Conversely, the solvent evaporation rate is strongly
decreased when employing the meniscus technique \cite{PaMo07}, i.e.,
by depositing a drop of solution on a Teflon ring that is wetted by
the solvent. This allows for a better control of the process and enables the use
of contrast-enhanced microscopy to observe the dewetting process in situ
\cite{Paul08}. All pattern formation is confined to the region of the
receding contact line of toluene, silicon and air.  With both
techniques one may find mono-modal or bi-modal polygonal networks
 \cite{MTB02}, labyrinthine spinodal structures, or branched
patterns (see Fig.~\ref{fig:finger}). The meniscus technique allows
for the study of branched structures in a more controlled manner.
The work in Ref.\ \cite{Paul08} indicates that fingering strongly depends on
the interaction strength of the particles, i.e., on the chain length
of the thiol molecules coating the gold cores.  For short chains
(C$_5$ and C$_8$) no formation of branched structures is observed. At
similar concentrations, well-developed branched structures are formed
for longer chains (C$_{10}$ and C$_{12}$).  For even longer chains
(C$_{14}$), however, one again finds less branching. It also
depends on the amount of excess thiol in the solvent (for details see Ref.\
\cite{Paul08}).

When following the evolution of the branched patterns in situ (see the
complementary video material of Ref.\ \cite{Paul08}), one clearly observes
that different processes occur on different lenght scales. First, a
macroscopic dewetting front recedes, leaving behind a seemingly dry
substrate. The macroscopic front can be transversely unstable
resulting in large-scale ($>100\mu$m) strongly anisotropic fingered
structures. For fronts that move relatively quickly these macroscopic
structures cover all the available substrate. However, when at a later
stage the macroscopic front becomes slower, those fingers become
scarce and `macroscopic fingering' finally ceases. At this stage it is
possible to appreciate that the seemingly dry region left behind by
the front is not at all dry, but covered by an ultrathin `postcursor'
film that is itself unstable.  The thickness of this film is similar to the
size of the nanoparticles.  At a certain distance from the
macroscopic front, the ultrathin film starts to evolve a locally
isotropic pattern of holes. The holes themselves grow in an unstable
manner resulting in an array of isotropically branched structures as
shown, e.g., above in Fig.~\ref{fig:finger}. This indicates that at
least some of the patterns described in the literature may
have arisen from processes in similar ultrathin `postcursor' films.

The existence of the ultrathin `postcursor' film is an experimental
finding that can be drawn on when choosing a theoretical approach to
account for the pattern formation (see
below).  Note however, that at the moment there exists no explanation
for its existence.  A possible hypothesis is that the substrate
strongly attracts the nanoparticles.  As a result they form a dense
suspension layer having a thickness roughly equal to the
diameter of the nanoparticles.  The observed mesoscopic dewetting front then actually
correspond to an autophobic dewetting of a low concentration
suspension from the higher concentration suspension on the surface of the substrate.

\section{Modelling approaches}
\label{sec:theo}

Models of dewetting thin films of pure liquids or polymers are often
based on thin film hydrodynamics.  Starting from the Stokes
equations, together with continuity and boundary conditions at the substrate and
free surface, one applies a long-wave approximation (assuming small
surface slopes and contact angles) \cite{ODB97,Thie07} and obtains a
non-linear evolution equation for the film thickness profile $h(x,y,t)$. In the case
of volatile liquids one finds \cite{SLT98,PKS99,LGP02,Pism04,Thie10}
\begin{equation}
\partial_t h \,=\,
\nabla\cdot\left[Q_{\mathrm{c}}\nabla\frac{\delta F}{\delta h}\right]
\,-\, Q_{\mathrm{e}}\frac{\delta F}{\delta h},
\mylab{eq:film}
\end{equation}
with the mobility functions $Q_{\mathrm{c}}(h)=h^3/3\eta\ge0$
(assuming Poiseuille flow in the film and no slip at the substrate; $\eta$ is the
dynamic viscosity) and $Q_{\mathrm{e}}\ge0$ for the convective and
evaporative part of the dynamics, respectively. $Q_{\mathrm{e}}$ is a
rate constant that can be obtained from gas kinetic theory or from
experiment \cite{LGP02}. Note that Eq.~(\ref{eq:film}) only applies if the pressure in the
vapour above the film is close to the saturation pressure.  For alternative
expressions that are used to describe the
non-conserved evaporative dynamics see, e.g., Refs.\
\cite{BBD88,OrBa99,PKS99,SREP01,KKS01,LGP02,BeMe06}.  Finally,
$\nabla=(\partial_x,\partial_y)$, and $\partial_t$, $\partial_x$ and
$\partial_y$ denote partial derivatives w.r.t.\ time and the
coordinates.

Focusing on the influence of capillarity and wettability only,
the energy functional $F[h]$ is given by
\begin{equation}
F[h]\,=\,\int dx \int dy \left[\frac{\gamma}{2}(\nabla h)^2 + f(h) - \mu h\right] 
\label{eq:en1}
\end{equation}
where $\gamma$ is the liquid-gas surface tension and $f(h)$ is a local
free energy term that describes the wettability of the surface.  Since $\mu$ corresponds to a chemical
potential, the term $\mu h$ may either bias the system towards the liquid or towards the gas
state. The variation of $F$ w.r.t.\ $h$ gives the pressure. It
contains the curvature (Laplace) pressure $-\gamma \Delta h$ and
the disjoining pressure
$\Pi(h)=-\partial_h f(h)$.  Many different forms for the latter are in
use (see, e.g., Refs.\ \cite{deGe85,TDS88,Isra92,Mitl93,ODB97,PiPo00,Thie07}).

For the present system a thin film description using
Eq.~(\ref{eq:film}) is not appropriate because the nanoparticles are not taken
into account.  However, under certain conditions one can augment
equation~(\ref{eq:film}) for the evolution of the film thickness by
coupling it to an equation for the evolution of the mean particle
concentration. The resulting model is able to describe the behaviour of an
evaporating solution on the meso- and macroscale.  Such an approach
is briefly discussed below in Section~\ref{sec:tfh}.  We should
expect such a model to describe the mesoscopic dewetting front discussed
above. However, the theory is less suited to a description of the
dewetting dynamics of the ultrathin postcursor film.

The dewetting of the ultrathin film of highly concentrated suspension
may be described by a discrete stochastic model such as, for instance,
a kinetic Monte Carlo (KMC) model based solely on
evaporation/condensation dynamics of the solvent and diffusion of the
solute \cite{RRGB03,Stan08,Vanc08}. The validity of this strong
assumption regarding the relevant transport processes can be confirmed
from an estimate based on Eq.~(\ref{eq:film}): The pressure $p=\delta
F/\delta h$ drives convection and evaporation.  The convective
mobility is proportional to $h^3$, i.e., it is large for thick films
but decreases strongly with reduced film thickness.  The
evaporative mobility, however, is a constant, implying that
evaporation will dominate below a certain (cross-over) thickness. For the
parameter values of Ref.\ \cite{LGP02} and a small contact angle
($\approx 0.01$), the cross-over thickness is in the range of 1-5
nanometers. This estimate justifies the neglect of convective
transport in a description of the postcursor film and may explain why
one has such good agreement between the experimentally observed
patterns and the patterns obtained from a purely two-dimensional
(single layer) kinetic Monte Carlo model \cite{RRGB03}.  We introduce
the KMC model below in Section~\ref{sec:kmc}.

In several respects, however, the kinetic Monte Carlo model is rather
simplistic, limiting its potential applications. For instance,
the thermodynamic chemical potential as well as any wetting
interaction of the solvent with the substrate are collected in a single
parameter -- an effective chemical potential. This implies that any
influence of a disjoining pressure is `smeared out' over the whole
system and that no distinction between the short- and the long-range parts of the
disjoining pressure is possible. It is furthermore based on the
assumption that evaporation/condensation is the dominant dynamic
process, but does not allow one to probe this assumption. In
Section~\ref{sec:ddft} we show how one may develop a dynamical density
functional theory (DDFT) that describes the system at a similar level to the KMC.
However, the DDFT may also be easily extended to include other effects
such as fluid diffusion, that the KMC does not incorporate.

\subsection{Kinetic Monte Carlo model}
\label{sec:kmc}

The kinetic Monte Carlo model for two-dimensional dewetting nanofluids
\cite{GeBr00} was first proposed in Ref.~\cite{RRGB03} and extended to
include next-nearest neighbour interactions in \cite{MBM04}.  The
two key assumptions used are: (i) the relevant processes can be
mapped on to a two-dimensional lattice gas model, thereby neglecting continuous
changes in the thickness of the evaporating film, and (ii) all
relevant dynamics results from diffusing nanoparticles and
evaporating/condensing solvent. 

The model builds on an Ising-type model for the liquid-gas phase
transition. The surface is divided up into a regular array of lattice
sites whose size is dictated by the nanoparticles.  One then considers
each lattice site to be occupied either by a nanoparticle, liquid or
vapour. This effectively maps the system onto a two-dimensional
two-component lattice gas having two fields $n$ and $l$. The resulting
three possible states of a cell are: liquid ($l=1, n=0$), nanoparticle
($l=0, n=1$), and vapour ($l=0, n=0$, i.e., cell empty). The energy of
an overall configuration is given by the hamiltonian
\begin{equation}
E\,=\,-\frac{\varepsilon_{nn}}{2}\sum_{<ij>} n_i n_j
\,-\,\frac{\varepsilon_{nl}}{2}\sum_{<ij>} n_i l_j
\,-\,\frac{\varepsilon_{ll}}{2}\sum_{<ij>} l_i l_j
\,-\,\mu\sum_{i} l_i
\mylab{eq:ham}
\end{equation}
where $\sum_{<ij>}$ denotes a sum over nearest neighbour pairs and
$\varepsilon_{ll}$, $\varepsilon_{nn}$ and $\varepsilon_{nl}$ are the
liquid-liquid, particle-particle and liquid-particle interaction
energies, respectively.  Fixing the three interaction strength
parameters $\varepsilon_{ll}$, $\varepsilon_{nn}$, $\varepsilon_{nl}$
and the effective chemical potential
$\mu$ determines the equilibrium state of the system.  We choose
$\varepsilon_{ll}$ as unit of energy -- i.e.\ we set
$\varepsilon_{ll}=1$.

The hamiltonian determines the equilibrium state and the energy
landscape of the system. However, as the system `dries in' during the
course of the solvent evaporation, the final nanoparticle
configurations do not necessarily represent equilibrium structures. This implies
that the system dynamics is of paramount importance. It is determined by the
possible Monte Carlo moves, their relative frequencies, and the
probabilities for their acceptance.  Two types of moves are allowed: (i)
evaporation/condensation of liquid and (ii) diffusion of nanoparticles
within the liquid. A mobility $M$ corresponds to the ratio of cycles
of particle and solvent moves and reflects the physical ratio of time
scales for evaporation and diffusion. A large mobility $M$ indicates
fast diffusion as compared to evaporation. A trial move is accepted
with the probability $p_{\mathrm{acc}} = \min[1,\exp(-\Delta E/kT)]$
where $k$ is the Boltzmann constant, $T$ the temperature and $\Delta
E$ is the change in energy resulting from the potential move.
Note that particles are only allowed to move into wet areas of the
substrate, i.e., onto cells with $l=1$. This models zero diffusivity
of the particles on a dry substrate. The replaced liquid fills the
site left by the nanoparticle.

\begin{figure}[htbp]
\centering
\includegraphics[width=0.8\hsize]{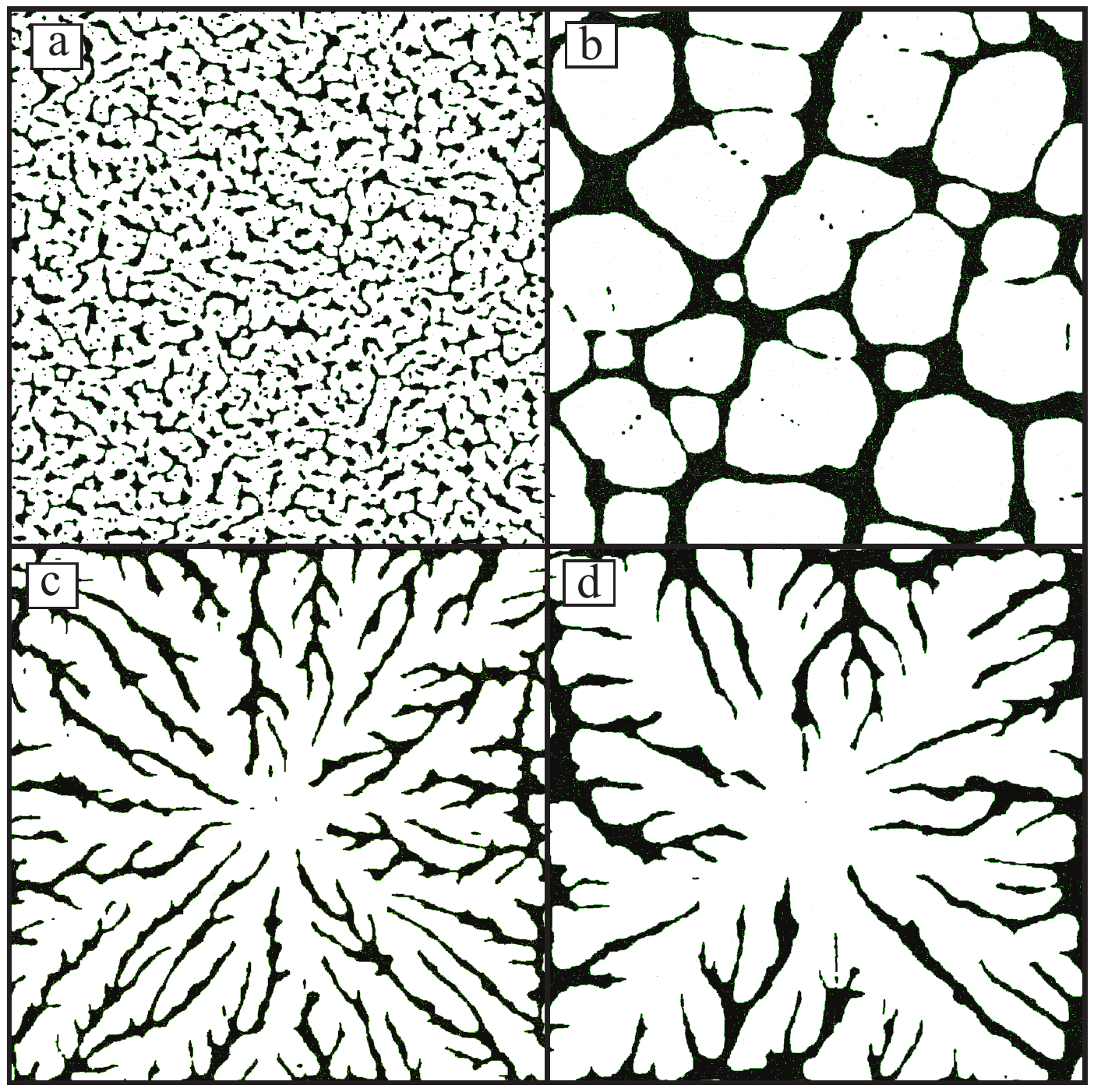}
\caption{Typical KMC results for the final dried-in nanoparticle structures resulting from
  the evaporative dewetting processes of nanoparticle solutions
  (nanofluids) in the case of (a) a spinodal-like process at
  $\mu=-2.55$, (b) nucleation and growth of holes at $\mu=-2.3$, (c)
  unstable fronts at $\mu=-2.3$ and low mobility $M = 5$,
  and (d) unstable fronts at $\mu=-2.3$ and medium mobility $M =
  10$. The starting configuration in (a) and (b) is a homogeneous liquid film
  with uniformly distributed particles whereas in (c) and (d) a hole at
  the center is nucleated `by hand'.  The remaining parameters are
  (a,b) $M = 50$, $\epsilon_{nl} = 2.0$, $\epsilon_{nn} = 1.5$, $\rho_n^{av}
  = 0.2$, $kT = 0.3$, MC steps$=500$, domain size $1200\times1200$;
  (c,d) $\varepsilon_{nn}=2.0$, $\epsilon_{nl} = 1.5$, $\rho_n^{av} = 0.2$,
  $kT = 0.2$, MC steps$=3000$, domain size $1200\times1200$. Lattice
  sites occupied by particles
  are coloured black, and the empty sites are coloured white. }
\mylab{fig:nanopatt}
\end{figure}

Without nanoparticles, the behaviour of the model is well known as it
reduces to the classical two-dimensional Ising model
\cite{Onsa1944}. For $kT<kT_c\approx0.567$ liquid and vapour
coexist when $\mu=\mu_{\rm coex}=-2$.  For $\mu>-2$ [$\mu<-2$] eventually
the liquid [vapour] dominates. A straight liquid-gas interface will recede
[advance] for $\mu<-2$ [$\mu>-2$], i.e.\ one finds evaporative
dewetting [wetting] fronts.  If one starts, however, with a
substrate covered homogeneously by the liquid, for $\mu<-2$ the film will dewet
via a nucleation or spinodal-like process. If the nanoparticles are
present, they form dried-in structures when all the liquid
evaporates. The final structures do not normally change any further -- at least on
short time scales.  However, if the liquid wets the particles (i.e.\ is attracted to the
particles), over long times there might be a coarsening of the structures, facilitated
by the adsorbed liquid. The dried-in patterns depend
on the particular pathway taken by the evaporative dewetting process. They range
from labyrinthine to polygonal network structures or holes in a dense
particle layer. Some typical patterns are displayed in
Fig.~\ref{fig:nanopatt}, for cases when the average surface coverage of the
nanoparticles $\rho_n^{av}=0.2$. Panels (a) and (b) result from a spinodal-like
and nucleation and growth process, respectively. At first sight they
look very similar to the patterns seen for the pure solvent and one
might argue that the particles solely act as passive tracers and
preserve the transient volatile dewetting structures of the
solvent. This was suggested in Refs.~\cite{Mert97,Mert98,TMP98} for
dewetting collagen solutions. However, panels (c) and (d) indicate
that the particles may at times play a rather more significant role.  When
the diffusion of the particles is slow, the evaporative dewetting fronts become
transversely unstable and may result in strongly
ramified patterns. This instability is caused by the nanoparticles.
The lower their mobility, the stronger the fingering effect, i.e., there are more
fingers in (c) than in (d) because in the latter the mobility is larger.

The front instability is intriguing as it results in strongly branched
structures. As the dewetting front moves, new branches are
continuously created and existing branches merge at the moving contact
line.  However, the mean finger number in the streamwise direction
of the resulting ramified pattern is a constant.  This behaviour is in contrast to the front
instabilities found for dewetting polymers which only result in fingers
without side-branches \cite{Reit93b} or fields of droplets left behind
\cite{ReSh01}.

A quantitative analysis shows that the mean number of fingers depends
only very weakly on the average concentration of the nanoparticles
$\rho_n^{av}$; only the mean finger width increases with increasing
concentration. However, decreasing the mobility (i.e., decreasing the
diffusivity of the particles) leads to a much denser finger pattern
and also causes the front instability to appear at an earlier stage,
i.e., when the front instability is in its initial linear regime, it
has a higher growth rate and a smaller characteristic wavelength
(cf.~Fig.~\ref{fig:nanopatt}(c) and (d)). Decreasing the effective
chemical potential (increasing its absolute value) has a similar but
less strong effect.  For details see \cite{Vanc08}.  These findings
lead to the conclusion that the determining factor for the front
instability is the ratio of the time-scales of the different transport
processes. In particular, the front becomes more unstable when the
velocity of the dewetting front increases as compared to the mean
diffusion velocity of the nanoparticles.

If the particle diffusivity is low, the front `collects' the
particles, resulting in a build up of the particles at the front that
itself is slowed down.  This makes the front unstable and any fluctuation
along the front will trigger a transverse instability that results in
an evolving fingering pattern. This happens even when the
particle-liquid and particle-particle attractive interactions do not
favour clustering (i.e.\ demixing of the liquid and the
nanoparticles).  In this regime, the instability is a purely dynamic
effect and energetics plays no role in determining the number of
fingers.  We call this the `transport regime'. 

To illustrate the influence of energetics (characterized by the
interaction parameters $\varepsilon_{ij}$) on fingering in
Fig.~\ref{fig:fingers-enl} we display the dependence of the mean
finger number on particle-liquid interaction strength
$\varepsilon_{nl}$.  For $\varepsilon_{nl}\ge1.5$ the mean finger
number $<f>$ is nearly constant; this is the transport
regime. However, on decreasing $\varepsilon_{nl}$ below 1.5, we
observe a marked increase in the value of $<f>$, indicating that
energy plays an important role in determining the number of fingers in
this regime.  In this parameter range, demixing of particles and
liquid occurs at the moving front and increases its transverse
instability.  In this `demixing regime', the wavelength of the
fingering instability is determined by the dynamics \textit{and} the
energetics of the system. Decreasing $\varepsilon_{nl}$ further (below
$1.4$ in Fig.~\ref{fig:fingers-enl}) one first observes in regime
(iii) a slight decrease in the average finger number. This is a
geometric effect resulting from our one-dimensional finger counting
routine: The fingers increasingly break up and the dried-in pattern
looks progressively isotropic. In regime (iv), the measure $\langle
f\rangle$ does not represent a finger number but instead indicates a
decrease in the typical distance between particle clusters resulting
from the demixing process that occurs already in the bulk liquid and
is not related to the front instability at all. Note that one finds a
similar sequence of regimes (i) to (iv) when increasing the
particle-particle interaction strengths for fixed $\varepsilon_{nl}$
(see Ref.~\cite{Vanc08}) for further details.

\begin{figure}[htbp]
\includegraphics[width=0.7\hsize]{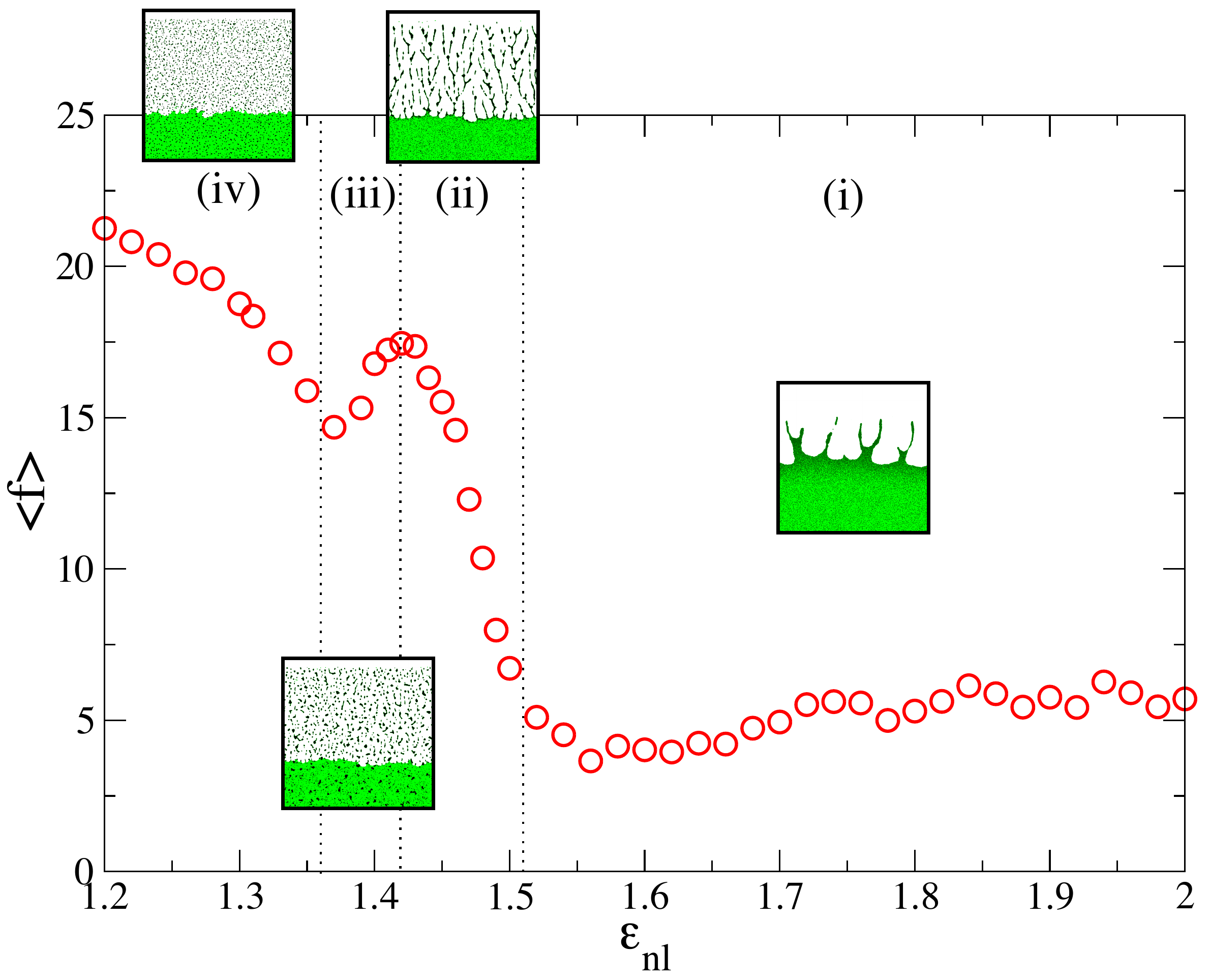}
\caption{(Colour online) Dependence of the mean finger number left
  behind by the unstable dewetting front on the particle-liquid
  interaction strength $\varepsilon_{nl}$. The regions marked (i) to
  (iv) are discussed in the main text. The insets display typical
  snapshots obtained in the four different regions. Particles are
  black, liquid is grey (green online) and the empty substrate is
  white. The remaining parameters are $kT = 0.2$, $M = 20$, $\mu =
  -2.2$, $\rho_n^{av} = 0.1$, $\epsilon_{nn} = 2.0$, domain size
  $1200\times1200$. For the insets, from left to right, $\epsilon_{nl} = 1.2,
  1.4, 1.45, 1.8$. }
\mylab{fig:fingers-enl}
\end{figure}

We note also that the fingering process may be viewed as
self-optimising the front motion -- i.e.\ the front keeps its average velocity
constant by expelling particles into the fingers. A similar effect exists
for dewetting polymer films \cite{ReSh01}, where liquid is expelled
from the growing moving rim which collects the dewetted polymer. There,
the surplus liquid is left on the surface as a droplet pattern.

The kinetic Monte Carlo model is a very useful tool that helps one to
understand the pattern formation in drying nanoparticle
suspensions. One has, however, to keep in mind the restrictions on the
model (see above). The purely two-dimensional character of the KMC was
extended to a `pseudo three-dimensional' one by making
the effective chemical potential dependent on the mean liquid
coverage \cite{Mart07}. As the latter is related to a mean film
thickness, this corresponds to the introduction of a `global'
thickness-dependent disjoining pressure into the evaporation term
without an explicit consideration of a film thickness. The amended
model can reproduce bimodal structures that are beyond the scope of
the purely two-dimensional model \cite{Mart07,Stan08}. Fully
three-dimensional models are also discussed in the literature
\cite{SHR05,YoRa06}.

\subsection{Dynamical Density Functional theory}
\label{sec:ddft}

The limitations of the kinetic Monte Carlo model introduced in the
previous Section are related to its character as a two-dimensional
lattice gas with only three states: gas, liquid or particle.  This
implies that (i) no liquid can be transported to a site on the surface
already filled with liquid, i.e., diffusion of the liquid can not be
incorporated in a sensible way and (ii) one is not able to distinguish between
the influence of the short- and the long-range parts of the interactions with
the substrate, as all such interactions are absorbed into the
effective chemical potential.

However, using dynamical density functional theory (DDFT) 
\cite{GPDM03,MaTa99,MaTa00,ArRa04,ArEv04,Mons08}
one can develop a model for the processes in the ultrathin postcursor
film without these limitations, although here we limit ourselves to
developing the theory at the level of the KMC and solely discuss how
to extend it to incorporate the influence of the liquid diffusion over the
surface. Such a DDFT model describes the coupled
dynamics of the density fields of the liquid $\rho_l$ and the
nanoparticles $\rho_n$.  The densities $\rho_l$ and $\rho_n$ are
defined as the probabilities of finding a given lattice site
on the surface to be occupied by a film of liquid or by a
nanoparticle, respectively.  Note that the probability densities correspond to
number densities as we use the lattice spacing $\sigma=1$ as our unit of length.

To develop the DDFT, one must first derive the underlying free energy
functional $F[\rho_l,\rho_n]$, and secondly, devise dynamical equations
for both density fields that account for the conserved and the non-conserved
aspects of their dynamics, i.e., transport and phase change
processes, respectively.

For a system governed by the hamiltonian (\ref{eq:ham}), we may
construct a mean-field (Bragg-Williams) approximation for the free
energy of the system \cite{ChLu97,GPDM03} which contains an entropic
contribution and contributions from the interactions between the
different species (nanoparticles and liquid).  The free energy is a
semi-grand free energy, since the liquid is treated grand canonically
(it is coupled to a reservoir with chemical potential $\mu$), whereas
the nanoparticles are treated in the canonical ensemble. The free
energy functional is first defined on the original KMC
lattice. However, after re-writing the interaction terms employing
gradient operators \cite{GPDM03} one finally obtains the free
energy functional for a continuous system
\begin{eqnarray}
F[\rho_l,\rho_n]&=&\int \dr \bigg[ f(\rho_l,\rho_n)
+\frac{\varepsilon_{ll}}{2}(\nabla \rho_l)^2 
+\frac{\varepsilon_{nn}  }{2}(\nabla \rho_n)^2
+\varepsilon_{nl}  (\nabla \rho_n)\cdot (\nabla \rho_l)
-\mu \rho_l \bigg],
\label{eq:free_energy}
\end{eqnarray}
where
\begin{eqnarray}
f(\rho_l,\rho_n)&=&kT[\rho_l\ln \rho_l+(1-\rho_l) \ln (1-\rho_l) ]\nonumber \\
&+&kT [\rho_n \ln \rho_n+(1-\rho_n) \ln (1-\rho_n) ]\nonumber \\
&-&2 \varepsilon_{ll} \rho_l^2-2 \varepsilon_{nn} \rho_n^2-4\varepsilon_{nl} \rho_n \rho_l.
\label{eq:f_loc}
\end{eqnarray}
%
Since the
liquid may evaporate from the surface into the vapour above the
surface, $\mu$ is the (true) chemical potential of this reservoir and
determines the rate of evaporation [condensation] from
[to] the surface.
Note that normally a free energy of the form in Eq.\
(\ref{eq:free_energy}) is obtained by making a gradient expansion of
the free energy functional of a continuous system
\cite{ChLu97}. However, here we have made the mapping from the free
energy of the lattice KMC system.

\begin{figure}[htb] 
\centering
\includegraphics[width=0.46\hsize]{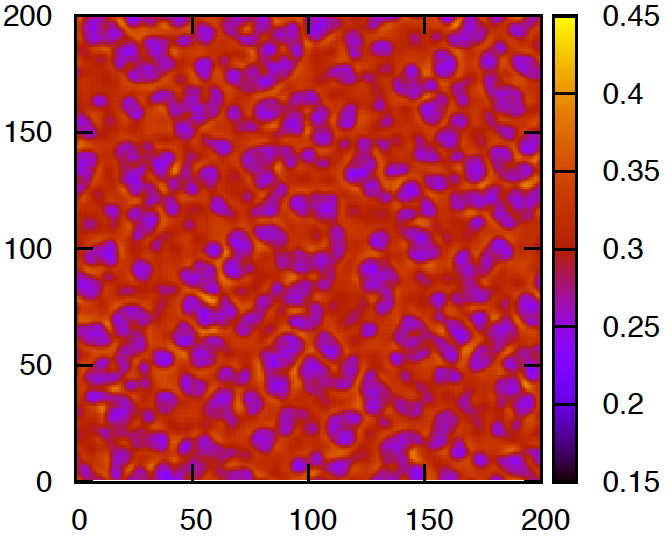}
\includegraphics[width=0.46\hsize]{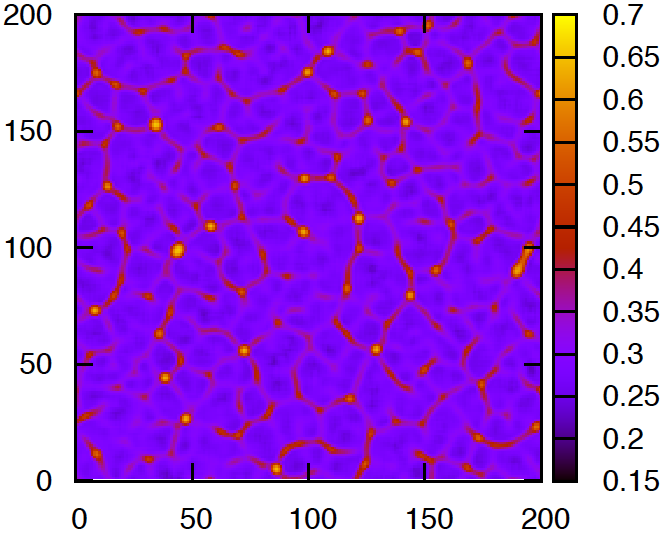}\\

\includegraphics[width=0.45\hsize]{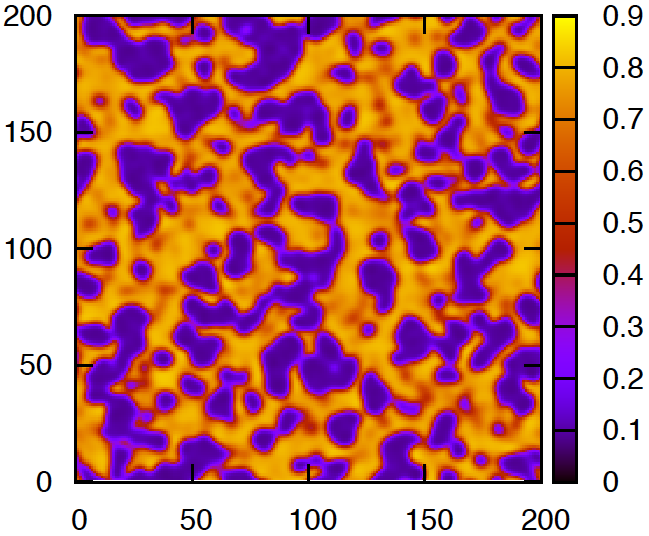}
\includegraphics[width=0.45\hsize]{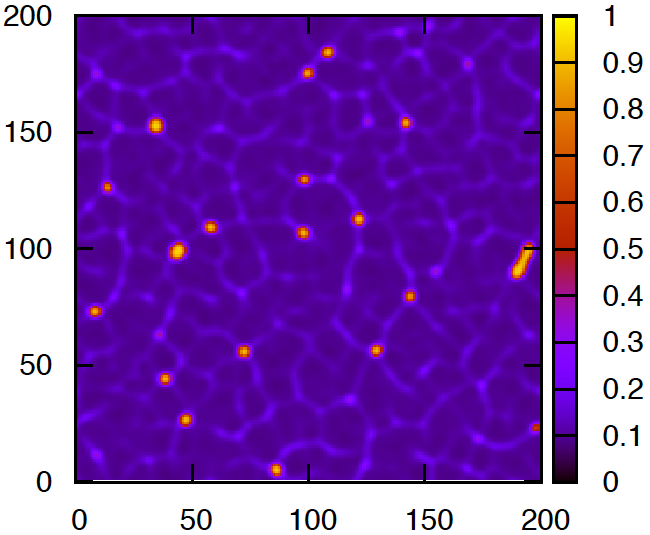}
\caption{(Colour online)
  Density profiles for the situation where the substrate is covered by
  nanoparticles with average density $\rho_n^{av}=0.3$. The top row
  are the nanoparticle density profiles and the bottom row are the
  corresponding liquid density profiles at the times $t/t_l=8$ (left)
  and $80$ (right), where $t_l=1/kT M_l^\mathrm{nc}\sigma^2$.  The
  parameters are $kT/\varepsilon_{ll}=0.8$,
  $\varepsilon_{nl}/\varepsilon_{ll}=0.6$, $\varepsilon_{nn}=0$,
  $\alpha=0.4M_l^\mathrm{nc}\sigma^4$, $M_l^\mathrm{c}=0$, $\rho_l(t=0)=0.9\pm\xi$ (where $\xi$ 
  represents white noise of amplitude 0.05) and
  $(\mu-\mu_\mathrm{coex})/kT=-0.88$, where the liquid exhibits spinodal
  decomposition-evaporation.
  }
\label{fig:ddft_spinodal}
\end{figure}

The chemical potential for the nanoparticles may be determined from
the functional derivative $\mu_n=\delta F[\rho_n,\rho_l]/\delta
\rho_n(\rr)$. In equilibrium it is constant throughout the system, but
it may vary spatially in a non-equilibrium system, i.e.,
$\mu_n=\mu_n(\rr,t)$.
We assume that the dynamics of the nanoparticles is governed by the
thermodynamic force $\nabla \mu_n$ -- i.e.\ that the nanoparticle current is
${\bf j}=-M_n \rho_n \nabla \mu_n$, where $M_n(\rho_l)$ is a mobility
coefficient that depends on the local density of the liquid. Combining
this expression for the current with the continuity equation,
we obtain the following evolution equation for the nanoparticle density profile
\begin{equation}
\frac{\partial \rho_n}{\partial t}=
\nabla \cdot \left[ M_n\rho_n \nabla \frac{\delta F[\rho_n,\rho_l]}{\delta \rho_n}\right].
\label{eq:DDFT_n}
\end{equation}
Note that this equation of motion may also be obtained by assuming
that the nanoparticles have over-damped stochastic equations of motion
\cite{MaTa00,ArRa04,ArEv04,Mons08}. Here, we assume that
$M_n(\rho_l)=\alpha \Theta_s(\rho_l-0.5)$, where $\Theta_s(x)$ is a
continuous function that switches smoothly from the value 0
to the value 1 at $x=0$ (i.e.\ it is essentially a smooth analogue of
the Heaviside function). This ensures that the
nanoparticles are immobile when the local liquid density is small (dry
substrate) and have a mobility coefficient $\alpha$ when $\rho_l$ is
high (wet substrate).

\begin{figure}[htb] 
\centering
\includegraphics[width=0.32\hsize]{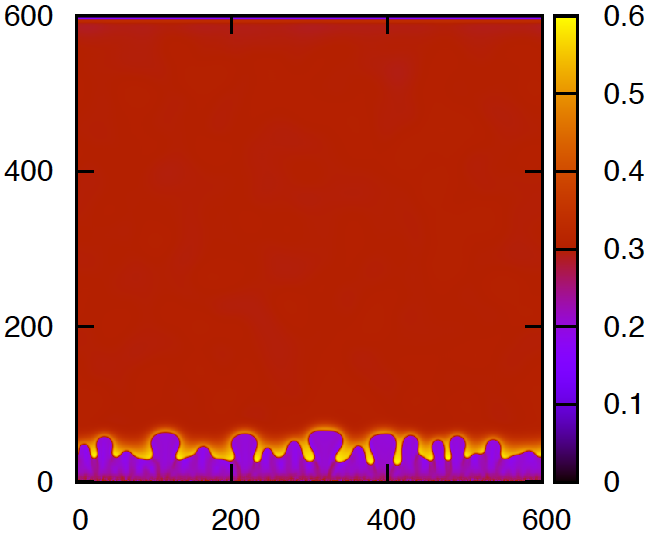}
\includegraphics[width=0.32\hsize]{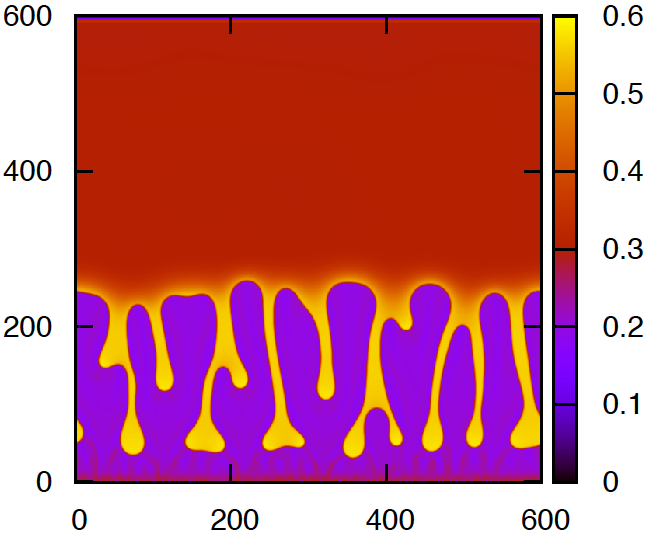}
\includegraphics[width=0.32\hsize]{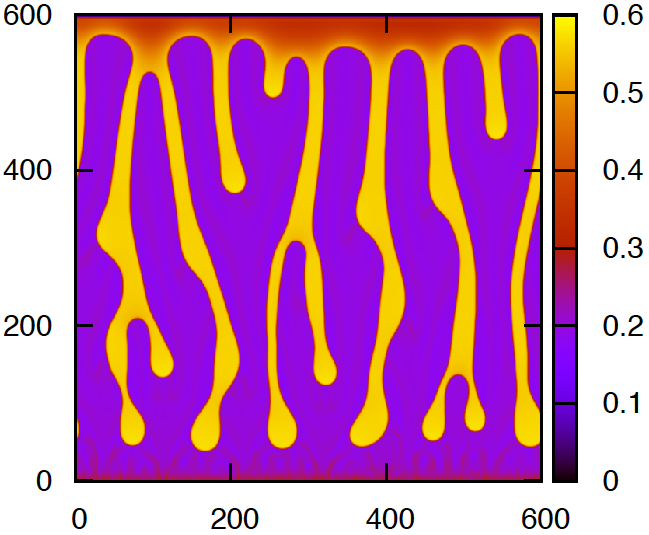}\\
\includegraphics[width=0.32\hsize]{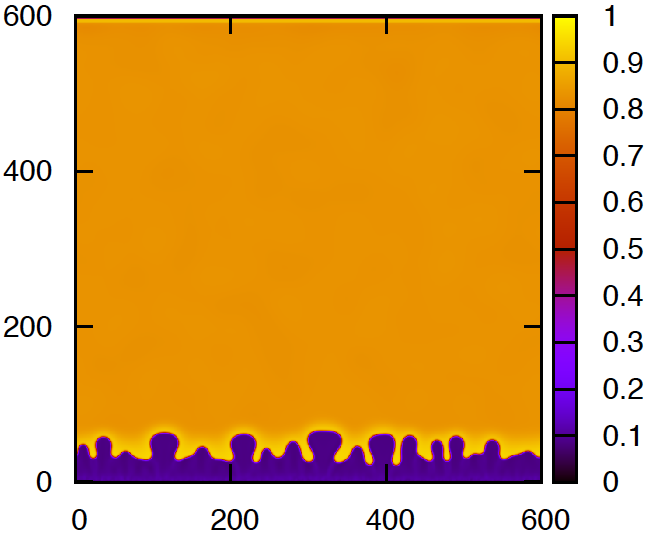}
\includegraphics[width=0.32\hsize]{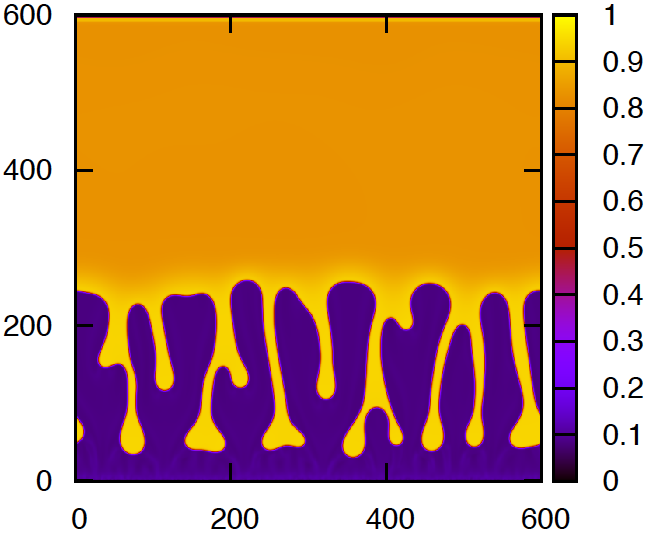}
\includegraphics[width=0.32\hsize]{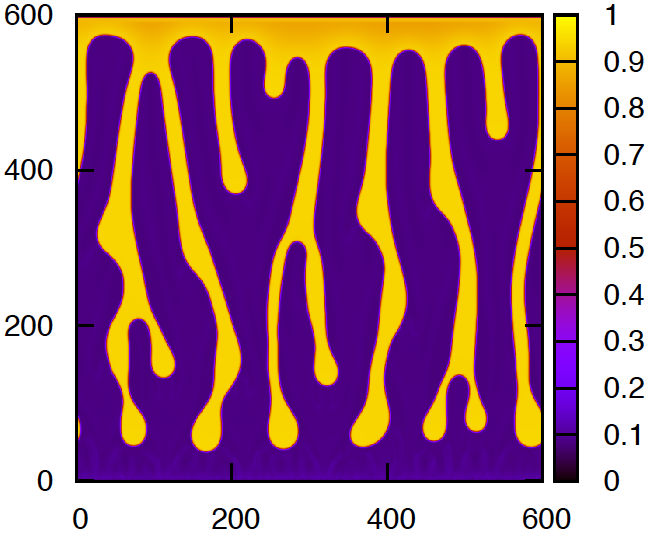}
\caption{(Colour online) Density profiles for the situation where the substrate is
  covered by nanoparticles with average density $\rho_n^{av}=0.3$ and
  with the liquid excluded from the region $y<0$.  The top row shows
  the nanoparticle density profiles and bottom row the corresponding
  liquid density profiles at the times $t/t_l=1000$ (left), $10000$
  (middle) and $30000$ (right), where
  $t_l=1/kT M_l^\mathrm{nc}\sigma^2$. The parameters are
  $kT/\varepsilon_{ll}=0.8$,
  $\varepsilon_{nl}/\varepsilon_{ll}=0.6$, $\varepsilon_{nn}=0$,
  $\alpha=0.2 M_l^\mathrm{nc}\sigma^4$, $M_l^\mathrm{c}=0$, $\rho_l(t=0)=0.9\pm\xi$ (where $\xi$ 
  represents white noise of amplitude 0.05)  and
  $(\mu-\mu_\mathrm{coex})/kT=-0.78$.}
\label{fig:ddft_fingers}
\end{figure}

For the evolution of the liquid density distribution we assume that
the liquid is able to evaporate from the surface into the vapour
(reservoir) above the surface (non-conserved dynamics) and may also
diffuse over the substrate (conserved dynamics). The conserved part is
treated along the lines developed above for the nanoparticles. For the
non-conserved part we assume a standard form \cite{Lang92}, i.e., the
change in time of $\rho_l$ is proportional to
$-(\mu_\mathrm{surf}(\rr,t)-\mu)=-\delta F[\rho_n,\rho_l]/\delta
\rho_l(\rr)$ where $\mu_\mathrm{surf}(\rr,t)$ is the local chemical
potential of the liquid at the point $\rr$ on the surface at time
$t$. This gives the evolution equation for the liquid density
\begin{equation}
\frac{\partial \rho_l}{\partial t}=\nabla \cdot 
\left[ M_l^\mathrm{c}\rho_l \nabla \frac{\delta F[\rho_n,\rho_l]}{\delta \rho_l}\right]
- M_l^\mathrm{nc} \frac{\delta F[\rho_n,\rho_l]}{\delta \rho_l},
\label{eq:DDFT_l}
\end{equation}
where we assume that the coefficients $M_l^\mathrm{c}$ and
$M_l^\mathrm{nc}$ are constants.

This theory allows us to study the time evolution of the
evaporating film of nanoparticle suspension without some of the restrictions
of the kinetic Monte Carlo model. Here, however, we illustrate its
application in similar parameter regimes as used above for the KMC.
We focus on two examples: (i) the spinodal dewetting of a initially
flat film of nanoparticle suspension characterised by constant
$\rho_l$ and $\rho_n$ (Fig.~\ref{fig:ddft_spinodal}); and (ii) the
retraction of a dewetting front that is unstable with respect to a
fingering instability (Fig.~\ref{fig:ddft_fingers}).

Fig.~\ref{fig:ddft_spinodal} presents two pairs of snapshots from a
purely evaporative dewetting process deep inside the
parameter region of the phase diagram where spinodal dewetting occurs.
For small times the film becomes unstable showing a
typical spinodal labyrinthine pattern with a typical wavelength. The
nanoparticles concentrate where the remaining liquid is
situated. However, they are `slow' in their reaction: when $\rho_l$
already takes values in the range 0.08 -- 0.83, the nanoparticle
concentration has only deviated by about 25\% from its initial value.
The film thins strongly forming many small holes. The competition for
space results in a fine-meshed polygonal network of nanoparticle
deposits. The concentration of particles is much higher at the network
nodes -- an effect that can not been seen within the KMC model. As the
particles attract the liquid there remains some liquid on the
substrate where the nanoparticles are.

Fig.~\ref{fig:ddft_fingers} gives snapshots of the evolution of a
fingering instability for a retracting dewetting front. At early times
the straight front shows a rather short-wave instability, about 16
wiggles can be seen. However, they are only a transient: the finger
pattern coarsens rapidly till only about 7 fingers remain. The
fingering then becomes stationary, i.e., just as in the KMC, the mean
finger number remains constant, although new branches are continuously
created and old branches join each other. In general, the results on
fingering agree well with results obtained using the KMC model
\cite{Vanc08}. From this we conclude that jamming of discrete
particles is not a necessary factor for causing the instability, since the fingering is
seen here in a continuum model with a diffusion constant that is independent
of the nanoparticle concentration.  The DDFT is better suited than the KMC
for investigations of the early instability stages: they are more easy to
discern without the discrete background noise of the KMC. Furthermore, one
may perform a linear stability analysis of the one-dimensional
undisturbed streamwise front profiles with respect to transverse
perturbations (in analogy to the approach used in Refs.\
\cite{SpHo96,ThKn03,MuWa05}).

\subsection{Thin film hydrodynamics}
\label{sec:tfh}
The previous two sections focused on two approaches to describe the
experimentally observed patterning dynamics in the ultrathin
postcursor film left behind by a mesoscopic receding dewetting front.
Although both the kinetic Monte Carlo model and the dynamical
density functional theory are able to describe well the processes in
the ultrathin film, they can not be employed to describe mesoscale
hydrodynamics.  A relatively simple model for the latter can be
derived in the framework of a long-wave or lubrication equation
\cite{ODB97,Thie07}. We will illustrate here the approach by
considering an isothermal situation where the nanoparticles are not
surface active, i.e., they do not act as
surfactants. For a model incorporating the effects of latent heat
generation and surface-active particles resulting in thermal and
solutal Marangoni stresses, see Ref.\ \cite{WCM03}.  A description of
spreading particle solutions incorporating a structural
disjoining pressure has also been considered \cite{MCS07}.
For related work on particle-laden
film flow on an incline see Refs.\ \cite{ZDBH05,CBH08}.

One starts from the Stokes equations, together with continuity, no-slip boundary
conditions at the substrate and force equilibria at the free surface,
and applies a long-wave approximation.  Under the assumption that
concentrations equilibrate rapidly over the film thickness, we obtain
coupled non-linear evolution equations for the film thickness profile
$h(x,t)$ and the amount of nanoparticles per unit length $h_p=\phi h$,
where $\phi$ is the volume concentration of the nanoparticles. Note, that $h_p$ corresponds
to the local thickness of the nanoparticle layer when all the solvent is
evaporated.  The resulting evolution equation for the film thickness
is Eq.\ (\ref{eq:film}) above and focusing on the influence of
%
%
particle-independent capillarity and wettability only, the energy
functional $F[h]$ is given by Eq.~(\ref{eq:en1}) above.  Note that the
viscosity $\eta$ depends on the particle concentration.
Following Refs.\ \cite{WCM03,MCS07,CBH08,CMS09}
we use the Quemada law for dense suspensions \cite{Quem77,QuBe02,StPo05}
\begin{equation}
\eta(\phi)=\eta_0\,\left(1-\frac{\phi}{\phi_c}\right)^{-2}
\mylab{eq:visc}
\end{equation}
where $\phi_c=0.64$ corresponds to random close packing of
spherical particles.  For the nanoparticle volume per length $h_p=\phi
h$ one obtains the following evolution equation:
\begin{equation}
\partial_t (\phi h) \,=\,\nabla\cdot \left[\phi Q_{\mathrm{c}}\nabla\frac{\delta F}{\delta h}\right]
+ \nabla\cdot \left[D(\phi) h \nabla \phi\right],
\mylab{eq:conc}
\end{equation}
where the particle concentration dependent diffusion coefficient
$D(\phi)$ is related to the viscosity by the Einstein relation
$D(\phi)=kT/6\pi R\eta(\phi)$, where $R$ is the radius of
the nanoparticles \cite{Dhon96}.

\begin{figure}[htb]
\centering
\includegraphics[width=0.8\hsize]{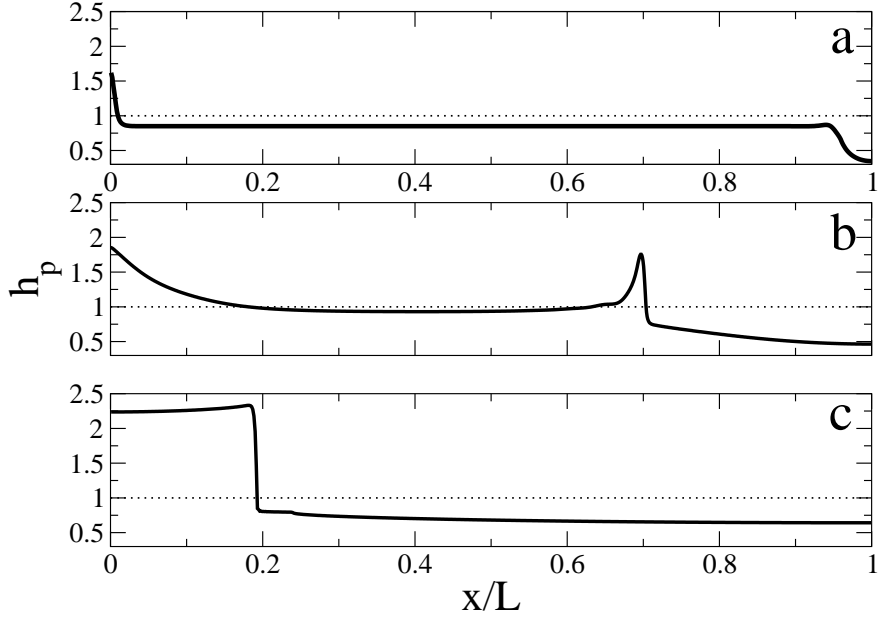}
\caption{Profiles of the final dried-in nanoparticle layer for the
  dewetting of a suspension of nanoparticles in a volatile solvent
  that partially wets the substrate for (a) high ($\Omega=10^{-3}$),
  (b) medium ($\Omega=2\times10^{-6}$) and (c) low
  ($\Omega=0.78\times10^{-8}$) evaporation rates, for the case when
  $\chi=H/l_0=1.09$, the lateral length scale is
  $\ell=\sqrt{\gamma/\kappa}H$ with $\kappa=(S_p/l_0)\exp(d_0/l_0)H$
  being an energy scale related to wettability and the vertical length
  scale is $H=\sqrt{2S_{LW}/\kappa} d_0$.  The remaining dimensionless
  parameters are the evaporation number $\Omega= Q_e\eta_0\ell^2/H^3$,
  the diffusion number $\Gamma=D(0)\eta_0/H\kappa=10^{-4}$ and the
  dimensionless chemical potential $M=H\mu/\kappa=-0.0035$. The system
  size is $L=19500\ell$. Film thickness and $h_p$ in the plots are
  scaled by the precursor film thickness.  }
\label{fig:driedin}
\end{figure}

We illustrate results obtained employing this thin film theory using
the single example of a receding dewetting front for a partially
wetting film.  We use the disjoining pressure and material constants
for the liquid considered in Ref.~\cite{LGP02}, where the evaporative
and convective dewetting of a film of volatile liquid is studied. We
add, however, the nanoparticles to the system. The expression that we
employ for the local free energy term in Eq.\ (\ref{eq:en1}) is:
\begin{equation}
f(h)=\frac{S_{LW}d_0^2}{h^2}+S_P \exp\left(\frac{d_0-h}{l_0}\right),
\label{eq:local_f}
\end{equation}
where the parameters characterising the interaction between the liquid film
and the surface are the apolar and polar spreading coefficients $S_{LW}$
and $S_P$, respectively, the Debye length $l_0$ and the Born repulsion
length $d_0$ \cite{LGP02}.  The resulting disjoining pressure
$\Pi=-\partial_hf(h)$ allows for a stable precursor film
(thickness $h_\mathrm{precursor}$) and also has a second (larger)
thickness ($h_0$) that corresponds to a secondary minimum of the
underlying energy functional. See Refs.\ \cite{TVN01,TVNP01} for
studies of film and drop states for similar disjoining pressures. Our
results are calculated for a system where the profiles only vary in
one Cartesian direction ($x$), corresponding to a straight dewetting
front. However, our results may also be interpreted as applying to a
circular flat drop whose front remains circular throughout the
dewetting and evaporation process. In this case one should interprete
the coordinate $x$ as the distance from the centre of the circular
film.

We start with a film of height $h_0$ of finite length sitting on a
precursor film and assume that the film contains nanoparticles at constant
concentration $\phi_0$. The chosen parameter values ensure that the
film of thickness $h_0$ is linearly stable. As we do not incorporate
noise, no nucleation of additional holes can occur (even with noise the
probability would be extremely low). Without evaporation the film
dewets `classically' by a retraction of the initially step-like
front. After a short time, surface tension smoothes the profile of the
receding front and a capillary rim forms that collects all the
dewetted liquid. The front recedes until all liquid is collected in a
central drop. Since no liquid evaporates [$Q_{\mathrm{nc}}=0$ in
Eq.\ (\ref{eq:film})], the particle concentration does
not change during the process.

The situation changes when allowing for evaporation
($Q_{\mathrm{nc}}>0$). Now the front may retract by convection
\textit{and/or} evaporation. Evaporation leads to the possibility of a
strong increase in the particle concentration at the contact line as
evaporation is strongest there. Due to the strong nonlinear dependence
of the viscosity on the particle concentration, this may lead to a
dramatic decrease of the convective contribution to the front
velocity.  For moderate evaporation rates, this may result in a
(temporary) self-pinning of the front. Within the present basic model,
the process can (after complete dry-in) result in three different
basic deposition patterns: (i) for very fast evaporation rates, all
other processes occur over time scales that are much larger.  In
particular, the effects of convective redistribution of the liquid are
neglectable. As a result one finds that a nearly homogeneous film of
nanoparticles of thickness $h_p=\phi_0 h_0$ is deposited (see
Fig.~\ref{fig:driedin}(a)). Convection only results in the small heap
of material visible at the left hand side of
Fig.~\ref{fig:driedin}(a). The decrease in $h_p$ on the right side of
Fig.~\ref{fig:driedin}(a) arises due to the diffusion of particles to
the right of the initial front position; (ii) for very low evaporation
rates, the film dynamics is dominated by convective dewetting as this
process acts on a much shorter time scale than evaporation.  As a
result, all the liquid is collected into a drop before evaporation
slowly removes the remaining solvent. Under these conditions most of
the nanoparticles are deposited in a single heap (see
Fig.~\ref{fig:driedin}(c)). Depending on the diffusivity, the heap
might be highest at the centre or show a depression there; (iii) at
intermediate evaporation rates, one may observe the deposition of a
nanoparticle ring around a region with a nanoparticle film of much
lower height. At the centre deposition might increase again (see
Fig.~\ref{fig:driedin}(b)).

\begin{figure}[htb]
\centering
\includegraphics[width=0.8\hsize]{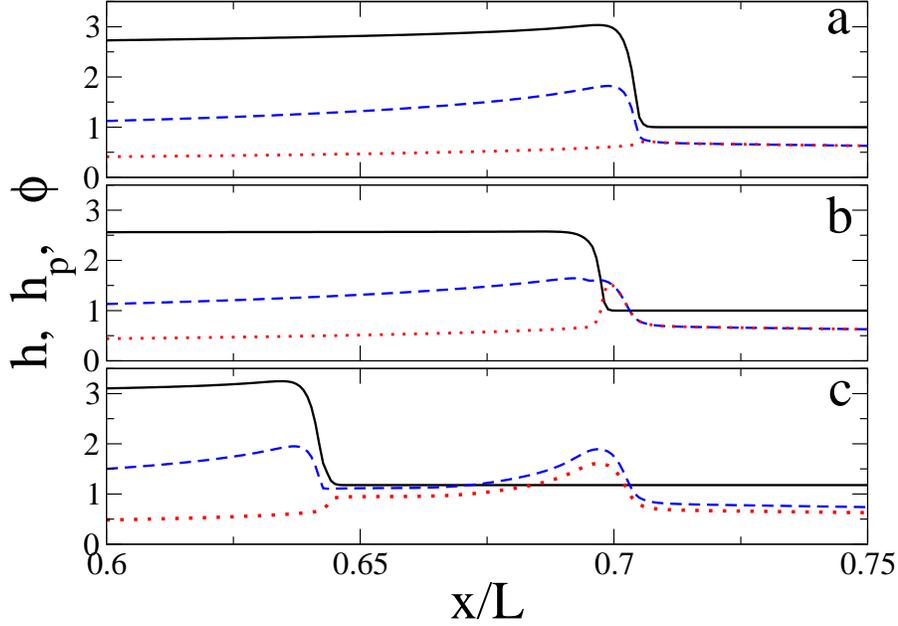}
\caption{(Colour online) A sequence of profiles during a dewetting
  process with competing evaporation and convection that leads to the
  dried-in ring structure of nanoparticles displayed in
  Fig.~\ref{fig:driedin}(b).  Profiles are at (a) before pinning ($t=
  0.08T$), (b) at self-pinning ($t= 0.13T$), and (c) after depinning
  ($t= 0.29T$), where $T=3\times10^{10}\tau$ with $\tau=\eta_0\gamma
  H/\kappa^2$ ($T$ is of order of 1s). The film thickness profiles $h$
  are the bold solid lines, the nanoparticle concentrations $\phi$ are
  the dotted lines and the nanoparticle layer height $h_p=h\phi$ are
  the dashed lines.  The remaining parameters and scalings are as in
  Fig.~\ref{fig:driedin}(b).}
\label{fig:ring}
\end{figure}

\begin{figure}[htb]
\includegraphics[width=0.49\hsize]{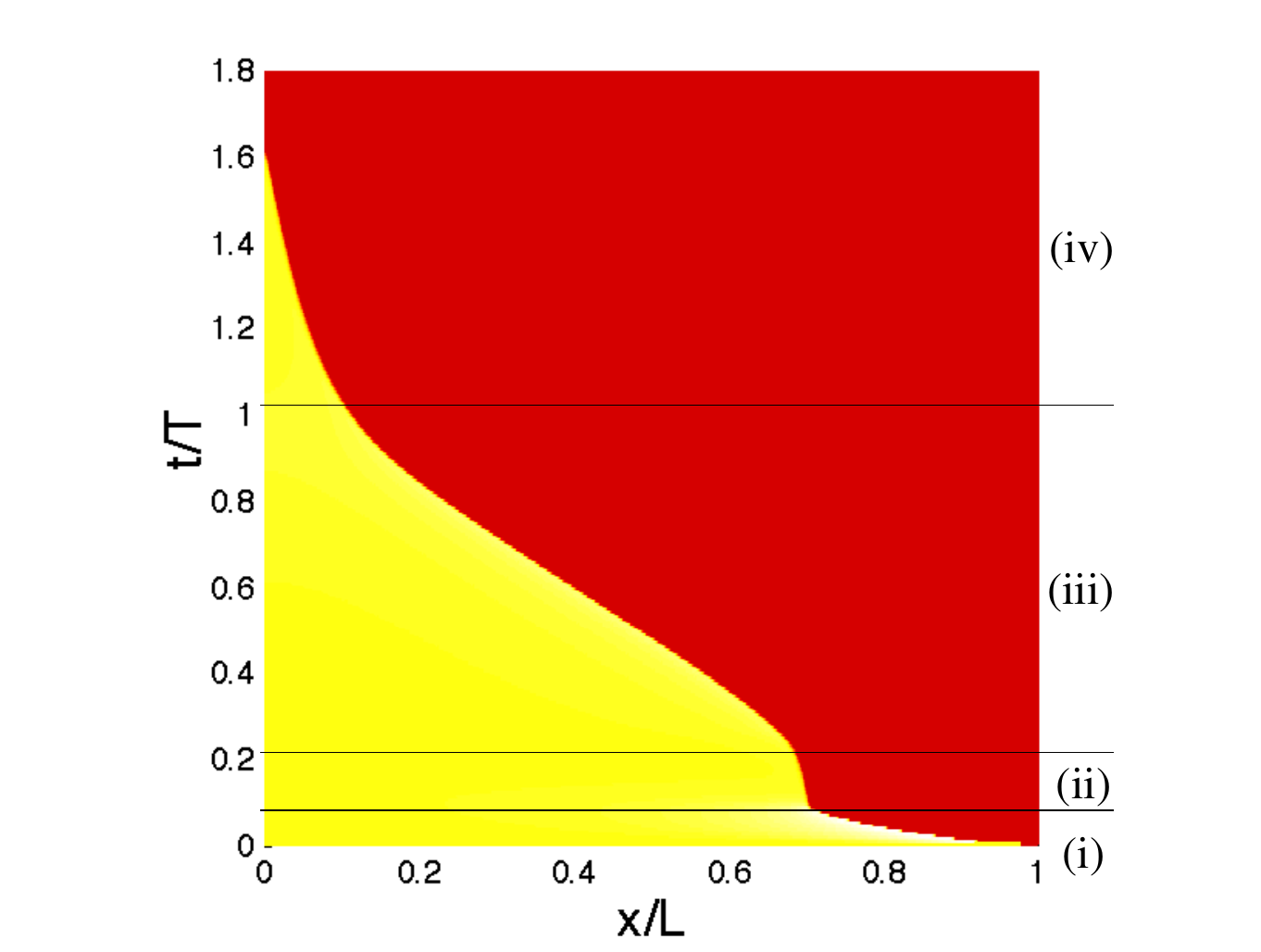}
\includegraphics[width=0.49\hsize]{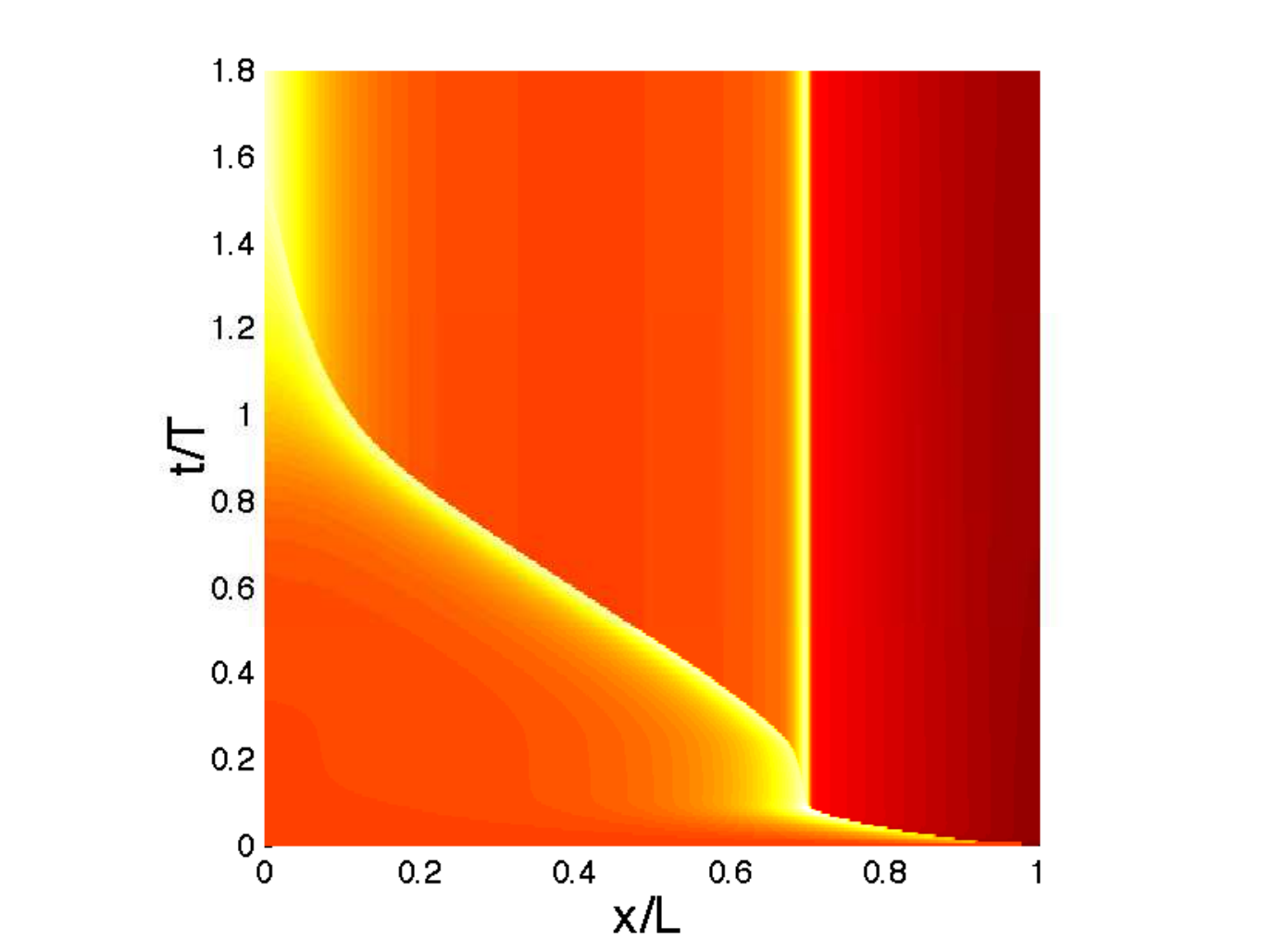}
%
\caption{(Colour online) Space-time plots are given for (left) the
  film thickness $h$ and (right) the nanoparticle layer height
  $h_p=h\phi$. The plot corresponds to the complete evolution
  resulting in the ring profile of Fig.~\ref{fig:driedin}(b). In both
  panels bright [dark] parts denote high [low] regions.  The prominent
  central dark-bright border in the left panel indicates the change of
  the position of the contact line in time. Over time, four regimes
  can be distinguished: (i) fast motion before pinning, (ii) nearly no
  front motion during self-pinning, (iii) slow motion after depinning,
  and (iv) final evaporation from the center.}
\label{fig:stplot}
\end{figure}

The most intriguing feature is the ring formation that has been
observed experimentally for suspensions of very different particle
sizes ranging from nanometers \cite{MMNP00,Deeg00,Deeg00b,GRPB04} to
hundreds of micrometers. Pinning of the contact line and thermal
Marangoni effects are often mentioned as necessary conditions for the
ring formation. The contact line pinning is often assumed to result
from substrate heterogeneities. Film height and concentration profiles
at various instants during the dewetting process are displayed in
Fig.~\ref{fig:ring}. The profiles are from before, at and after
self-pinning of the contact line. In Fig.~\ref{fig:stplot} we display
a space-time plot for the complete process. At first, the front
recedes in the same manner as when there is no evaporation, but now
driven by convection and evaporation. A small capillary rim forms that
collects all the dewetted liquid that does not evaporate. The particle
concentration slowly increases at the contact line
(Fig.~\ref{fig:ring}(a) and regime (i) in Fig.~\ref{fig:stplot}).  The
concentration increases further and when it approaches random
close packing $\phi_c$, the viscosity diverges and the front pins
itself. When pinned, further retraction only occurs through
evaporation (Fig.~\ref{fig:ring}(b) and regime (ii) in
Fig.~\ref{fig:stplot}).  The front eventually depins and starts to
move again, leaving a nanoparticle ring behind (Fig.~\ref{fig:ring}(c)
and regime (iii) in Fig.~\ref{fig:stplot}). However, the velocity is
not as large as at the beginning, owing to the fact that the mean
concentration of particles has increased. The remaining particles are
transported to the centre and are deposited there when the remaining
solvent evaporates (regime (iv) in Fig.~\ref{fig:stplot}).

The simple model used here shows, (i) that the contact line stops due
to self-pinning by the deposited particles and (ii) the Marangoni
effect is not necessary for the ring formation. The model can easily
be refined to account for solutal and/or thermal Marangoni effects
\cite{WCM03} but self-pinning should also be investigated further in the
simple case presented here.

\section{Conclusion}
\label{sec:conc}

We have discussed recent work on pattern formation processes in films
and drops of evaporating suspensions/solutions of polymers and
particles. After reviewing experiments on suspensions
of thiol-coated gold nanoparticles in toluene we have focused on the
modelling of the transport and phase change processes involved.  A
theoretical approach to the modelling of the hydrodynamics on the
mesoscale has been described as well as more microscopic models for the dynamics in
the observed nanoscopic `postcursor' film.
In particular, we have introduced (i) a microscopic kinetic
Monte Carlo model, (ii) a dynamical density functional theory and
(iii) a hydrodynamic thin film model.

The kinetic Monte Carlo model and the dynamical density functional
theory can both be used to investigate and understand the formation of
polygonal networks, spinodal and branched structures resulting from
the dewetting of an ultrathin `postcursor' film that remains behind
the mesoscopic dewetting front. They are, however, not capable of
describing the dynamical processes in a mesoscopic film. We have seen
that the KMC model is able to describe the interplay of solute
diffusion within the solvent and solvent evaporation/condensation.  It
also takes the liquid-liquid, liquid-particle and particle-particle
interactions into account and therefore allows us to distinguish
different regimes of the transverse (fingering) instability of the
evaporative dewetting front: a transport regime where the instability
is almost completely independent of the interaction strengths and a
demixing regime where particles and liquid demix at the receding front
thereby increasing its transverse instability.

The dynamical density functional theory describes the coupled dynamics
of the density fields of the liquid and the nanoparticles. In the form
described above (i.e.\ based on the two-dimensional hamiltonian
(\ref{eq:ham})) we obtain a simple theory that allows us to study the
time evolution of the evaporating ultrathin film and also to
investigate the influence of processes such as surface diffusion by
the liquid, which are not incorporated in the KMC model. However, it
is straightforward to extend the theory to consider a fully
three-dimensional fluid film, in which one can distinguish between
short- and long-range interactions of solvent and/or solute with the
substrate. We have, however, restricted the examples given here to
situations that can also be described using the KMC model. A further
exploration will be presented elsewhere.

Finally, we have discussed a simple thin film model for the
hydrodynamics on the mesoscale. It results from a long-wave
approximation and consists of coupled evolution equations for the film
thickness profile and the mean particle concentration. It has been
used to discuss the self-pinning of receding contact lines that is
related to the formation of rings of dried-in particles (coffee-stain
effect) that frequently occurs when films or drops of solutions or
suspensions dewet by the combined effects of convection and
evaporation.

One of the primary goals of researchers in this field, is the
search for simple-to-use techniques that allow one to produce
hierarchically structured functional layers for a wide range of
applications such as, e.g., organic solar cells \cite{Heie08}. This
means that the experiments advance very rapidly towards increasingly
complex systems.  For example, there have been investigations of the
influence of the phase behaviour on the drying of droplets of a
suspension of hard-sphere colloidal particles and non-adsorbing
polymer \cite{HGP02}, of the instabilities and the formation of drops
in evaporating thin films of binary solutions \cite{GPBR05} that may
lead to treelike patterns \cite{GRBP06b}, of effects of a secondary
phase separation on evaporation-induced pattern formation in polymer
films \cite{YNKA02}, and of the influence of an imposed flow on
decomposition and deposition processes in a sliding ridge of
evaporating solution of a binary polymer mixture \cite{Muel06} and of
the influence of rather fast evaporation \cite{Borm05,Borm05c}. These
complex experimental systems all represent systems of high practical
interest that the theories presented here are not (yet) able to
describe. Such experiments do, however, provide a strong motivation
for further work to extend the theories presented here, as well as to
develop new approaches.

Let us finally mention that several topics were entirely excluded from
our discussion here. First, we focused on a limited range of
descriptions and did, for instance, not mention lattice Boltzmann,
molecular dynamics or dissipative particle dynamics approaches that
may also be employed to describe fluid suspensions
\cite{Gibs99,StPa08,DKKA05,Krom06}. Second, we have only discussed
spatially homogeneous substrates. Patterned substrates are widely used
in dewetting experiments
\cite{Rock99,Sehg02,GeKr03,Mart07}. Theoretical descriptions are well
developed for the dewetting of films of pure non-volatile liquids on
such substrates
\cite{LeLi98,BDP99,KKS00,KKS01,BrLi02,BKTB02,TBBB03,Thie03}. However,
in the case of volatile liquids on heterogeneous substrates, much less
work has been done. A third topic that we did not touch upon are
possible continuum thin film approaches to demixing dewetting
suspensions. We believe it is feasible to extend the diffuse interface
theories such as model-H \cite{AMW98} to include the influence of
evaporation in dewetting nanoparticle suspensions.  For instance, such
models have already been adapted to describe demixing free surface
films of polymer blends \cite{TMF07,FNOG08,MaTh09}.

\acknowledgements
AJA and MJR gratefully acknowledge RCUK and EPSRC, respectively,
for financial support. We acknowledge support by the European Union
via the FP6 and FP7 Marie Curie schemes [Grants MRTN-CT-2004005728
(PATTERNS) and PITN-GA-2008-214919 (MULTIFLOW)].\\[2ex]

\end{document}